\documentclass[10pt,a4paper,twoside,onecolumn]{article}
\usepackage[english]{babel}
\usepackage{latexsym}
\usepackage{amssymb}
\usepackage[dvips]{graphicx}

\def\titlefont{\fontsize{12}{13}\bfseries\boldmath\selectfont\centering{}}
\def\authorfont{\footnotesize}

\let\affiliationfont\rhfont

\def\title#1{
    \vspace*{-14pt}
    \vskip 79pt
        {\centering{\titlefont #1\par}}%
    \vskip 1em
}

\def\author#1{\par
    {\centering{\authorfont#1}\par\vspace*{0.05in}}
}

\def\address#1{\par
    {\centering{\affiliationfont#1\par}}\par\vspace*{11pt}
}

\def\keywords#1{\par
    \vspace*{8pt}
    {\authorfont{\leftskip18pt\rightskip\leftskip
    \noindent{\it Keywords}\/:\ #1\par}}\vskip-12pt}
  \def\body{
\setcounter{footnote}{0}
\def\thefootnote{\alph{footnote}}
\def\@makefnmark{{$^{\rm \@thefnmark}$}}
}  
\def\bodymatter{\body}

\begin{document}

\title{Exclusive vector meson electroproduction @ CLAS} 

\author{A. Fradi}

\address{Univ Paris-Sud, Institut de Physique Nucl\'eaire d'Orsay\\
91405 Orsay,  France\\
$^*$E-mail: fradi@ipno.in2p3.fr\\}

\begin{abstract}
We present the results of exclusive electroproduction of vector mesons on the proton at CLAS.
We discuss the interpretation of these cross sections in terms of
$t$-channel Reggeon exchanges and in terms of Generalized Parton
Distributions (GPDs) formalism. 
\end{abstract}

\bodymatter
\section{Introduction}\label{aba:sec1}
The exclusive electroproduction of mesons on the nucleon is an important tool to better understand the
nucleon structure and, more generally, the transition between the low energy hadronic and high energy partonic
domains of the Quantum Chromodynamics (QCD) theory.
 The CEBAF (``Continous Electron Beam Accelerator Facility'') of the Jefferson Laboratory (JLab) at Newport News (USA) with
 its high intensity electron beam  and its large acceptance spectrometer CLAS \cite{clas} (CEBAF Large Acceptance Spectrometer) 
 offers a great opportunity to study the exclusive electroproduction of mesons,
 in particular those that have multi-particle decays.\\
 This proceeding presents the
 results for the exclusive electroproduction of the vector mesons $\rho^0$, $\omega$ and $\phi$ on the
 proton  at CLAS and provides a first (preliminary)
look at the $\rho^+$ channel. These results come mainly from two experiments: the e1-6  and the
 e1-dvcs experiments. For e1-6, the data were collected between October 2001 and January 2002. The beam energy was 5.754 GeV and the 
 integrated luminosity was 28.5 fb$^{-1}$. The e1-dvcs data were collected between March  and May 2005. The beam energy 
 was 5.75 GeV and the integrated  luminosity was $\sim 40$ fb$^{-1}$. The cross
 sections for the $\rho^0$, $\omega$ and $\phi$
 channels are published in
 refs. \cite{rho,omega,phi}. For the $\rho^+$ channel, it is the first ever measurement of its cross section.
The presented results are still preliminary and more details can be found in ref. \cite{mathese}.
 \section{Data analysis}      
\subsection{The reaction $e p \to e  p \rho^0$}
\begin{figure}[!h]
\begin{center}
\includegraphics[width=9cm]{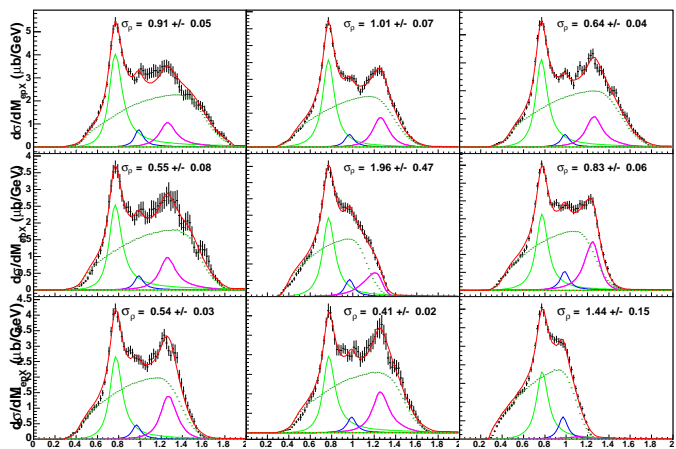}
\caption{Missing mass $M_X[e p X]$ for some $(Q^2,x_B)$ bins. See ref. \cite{rho} for the kinematics of
each bin.}\label{missigep}
\end{center}
\end{figure}

The final state  that  is analyzed is $e p \to e p \pi^+ \pi^-$.
The final state particles $e$, $p$ and $\pi^+$ were detected in CLAS and the exclusive process was identified by selecting 
a missing $\pi^-$ with the missing mass technique. Fig. \ref{missigep} shows the distributions of $M_{X}[e p X]$ for some $(Q^2,x_B)$ bins. 
One sees clearly the $\rho^0$ peak sitting on top of a background of a non-resonant two-pion continuum. One can also distinguish two
bumps at masses around 980 MeV and 1270 MeV corresponding, respectively, to the scalar $f_{0}$ and tensor $f_{2}$ mesons.\\
After weighting each event with the appropriate acceptance, we have used four  contributions to fit these spectra: three skewed Breit-Wigner distributions to describe
the $\rho^0$, $f_0$ and $f_2$ resonant structures and a parametrisation of the $M_{\pi^+\pi^-}$ 
non-resonant continuum determined  from simulations (mostly
$e p \to e p \pi^{+} \pi^{-}$ phase space). After the background subtraction procedure, total and differential cross 
sections of  the reaction  $e p \to e p \rho^0$ were extracted, by calculating for each $(Q^2,x_B)$ bin the area of the $\rho^0$ Breit-Wigner. 
A separation of the longitudinal cross sections from  the transverse ones was performed  by analysing the pion 
decay angles of the $\rho^0$ and relying on the SCHC ($s$-channel helicity conservation) 
which was checked experimentally.
\subsection{The reaction $e p \to e p \omega$}
\begin{figure}[!h]
\begin{center}
\includegraphics[height=3cm,width=5cm]{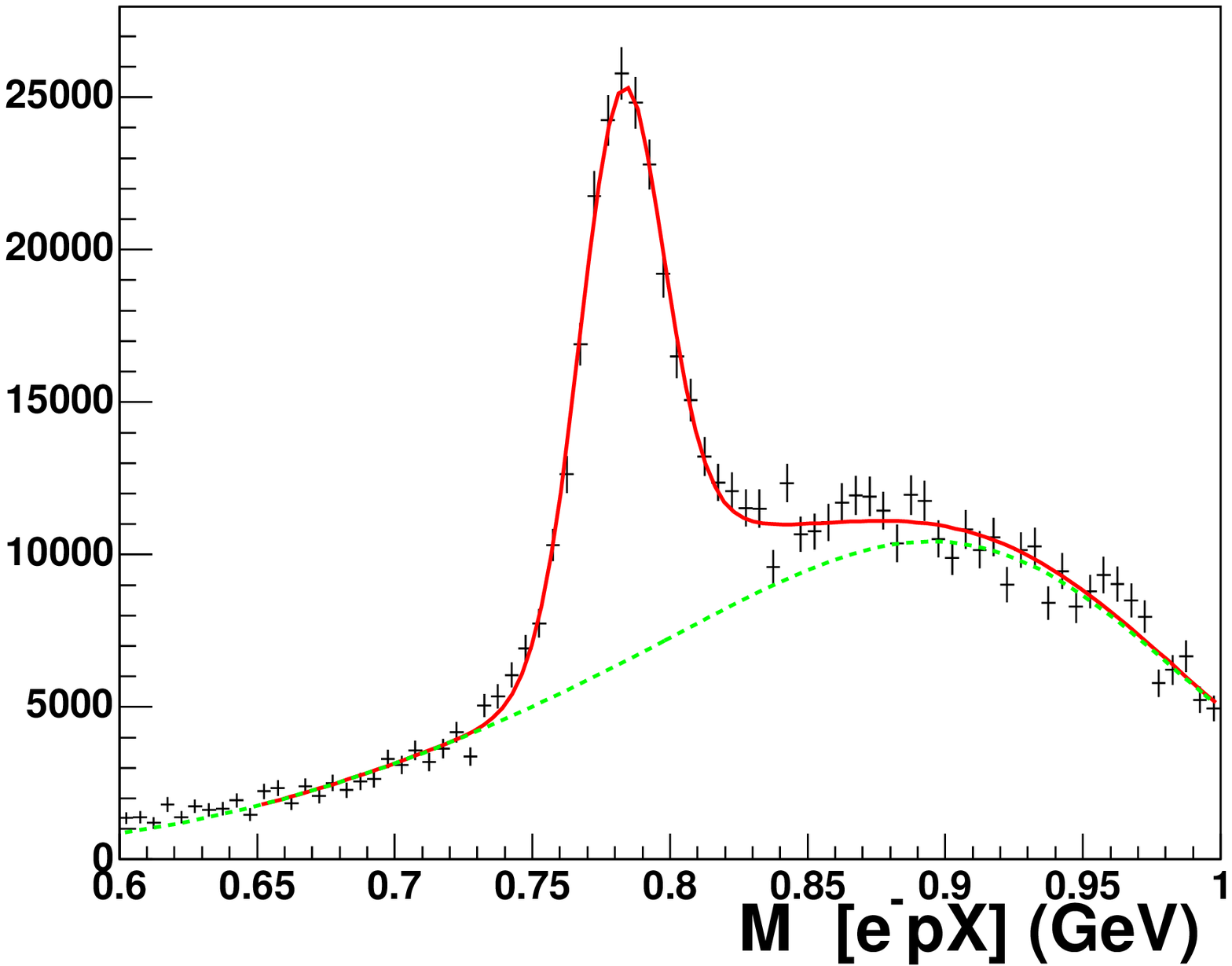}\hfill
\includegraphics[height=3cm,width=5cm]{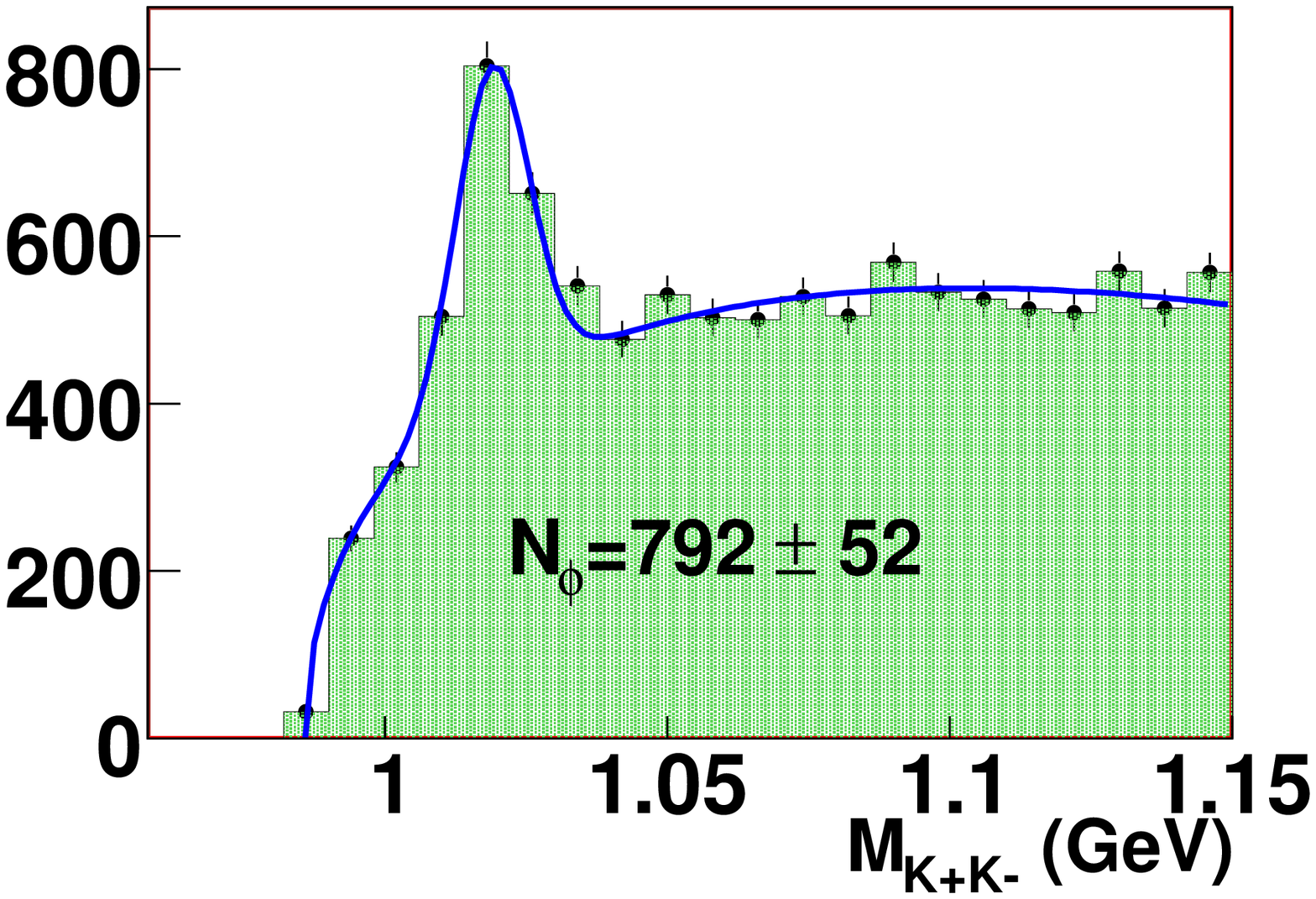}
\caption{Left: missing mass $M_{X}[epX]$ distribution for a particular ($Q^2,x_B$) bin. Right: invariant mass $M_{K^+ K^-}$ for the
whole data set.}\label{omegaphi}
\end{center}
\end{figure}

The final state that is analyzed is $e p \to e \pi^+ \pi^- \pi^0$. The final state particles $e$, $p$ and $\pi^+$ were detected in CLAS and the exclusive process was identified by 
 the missing mass $M_{X}[epX]$. Fig. \ref{omegaphi} (left) shows this acceptance-corrected missing mass distribution for a particular ($Q^2,x_B$) bin.
The $\omega$ peak is fitted with a skewed Gaussian shape (taking into account the experimental  resolution and radiative tail) and the background with a second-order polynomial.
 \subsection{The reaction $e p \to e p \phi$}
The final state that is analyzed is  $e p \to e p K^+ K^-$. The final state particles $e$, $p$ and $K^+$ were detected in CLAS and the exclusive process was identified by selecting 
a missing $K^-$ with the missing mass technique. Fig. \ref{omegaphi} (right) shows the invariant mass  $M_{K^+ K^-}$ for the entire data set with a
clear $\phi$ meson peak. This distribution is fitted with a  Gaussian plus a polynomial function for the background.
After weighting each event 
with the acceptance, the  background  was subtracted and the cross section of the exclusive electroproduction of $\phi$ was extracted.
\subsection{The reaction $e p \to e  n \rho^+$}
\begin{figure}[!h]
\begin{center}
\includegraphics[width=12cm]{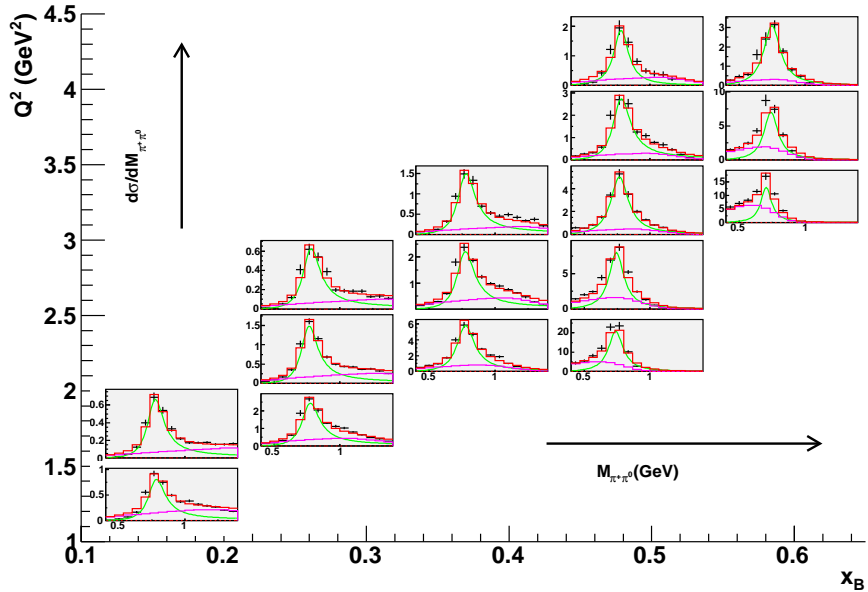}
\caption{$M_{\pi^+ \pi^0}$ acceptance-corrected distributions, showing fits for the background
subtraction. In black: experimental data; In green: Breit-Wigner of $\rho^+$; in purple: $M_{\pi^+ \pi^0}$ 
projection of the the non-resonant continuum $\gamma^* p \to n \pi^+ \pi^0$ reaction; in red: total fit result. In this proceeding, units are arbitrary for the vertical scale.} \label{fitQ2xB}
\end{center}
\end{figure}

The final state that is analyzed is $e p \to e n \pi^+ \pi^0$. The final state particles $e$, $\pi^+$ and $\pi^0$ were detected in CLAS and the exclusive process was identified by selecting 
a missing neutron  with the missing mass technique. Fig. \ref{fitQ2xB} shows the
acceptance-corrected invariant mass $M_{\pi^+ \pi^0}$  distributions for each
  $(Q^2,x_B)$ bin. One sees clearly the $\rho^{+}$ peaks around 775 MeV. They  are  however
  sitting on top of a non-negligeable non resonant two-pion background.  We have used two contributions to fit these spectra: a skewed Breit-Wigner distribution to describe
  the resonant structure of the $\rho^+$ and a parametrization
of the non-resonant two-pion continuum  determined from simulations (mostly $e p \to e n \pi^+ \pi^0$ phase space). 
In order to extract the total cross section of the reaction  $\gamma^{*}  p \to  n  \rho^{+}$, we have calculated for each 
    $(Q^2,x_B)$ bin the area of the Breit-Wigner (green curve in Fig. \ref{fitQ2xB}), the distributions being already normalized 
    and corrected by the acceptance.
\section{Discussion}\label{theory}
\begin{figure}
\includegraphics[height=3cm]{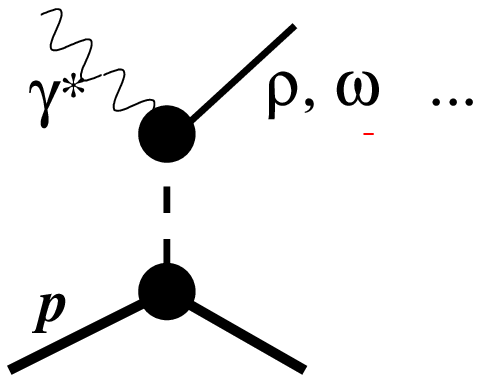}\hfill
\includegraphics[height=3cm]{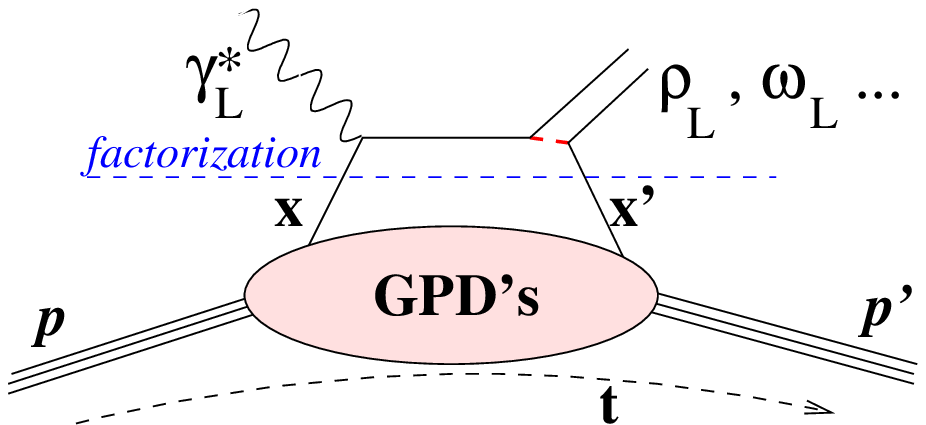}
\caption{Shematic representations of the Reggeon $t$-channel exchange (left) and of the handbag diagram (right) for exclusive vector
  meson electroproduction.}\label{reggeGPDs}
\end{figure}

 We now compare the extracted total and differential cross sections of exclusive
 electroproduction for these vector mesons with two theoretical approaches: on the one hand, the hadronic approach based on Regge
 theory and meson trajectory exchanges in the $t$-channel and on the other hand, the partonic approach based on the handbag diagram and GPDs. 
 These two approaches are illustrated in Fig. \ref{reggeGPDs}.
This comparison will be useful in order to  better understand the
 domain of validity of these two approaches and constrain their inputs.
   \subsection{The Regge ``hadronic'' approach}
\begin{figure}[h]
\begin{center}
\includegraphics[width=11cm]{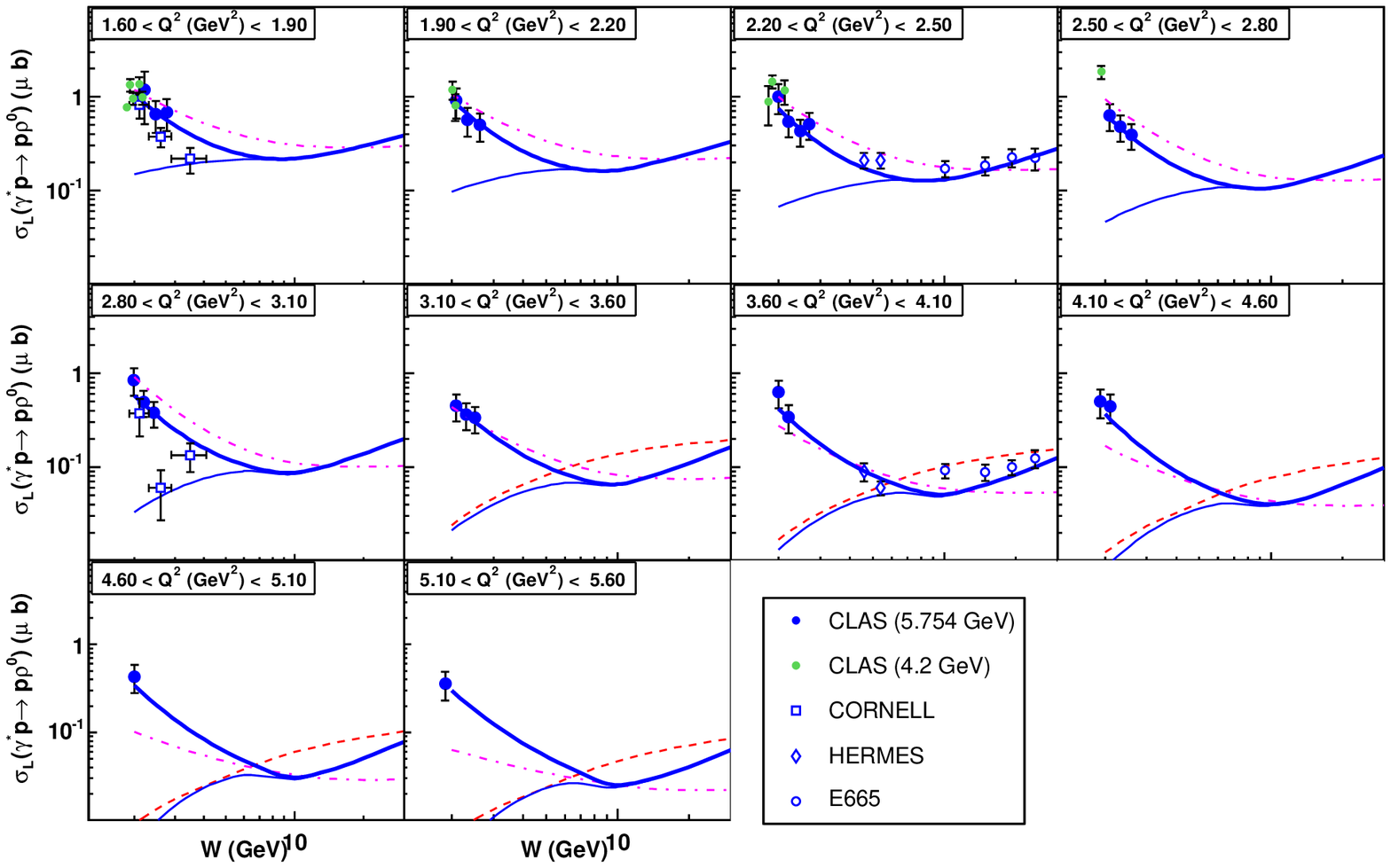}
\caption{Longitudinal cross section as a function of $W$ at fixed $Q^2$, for the reaction 
$\gamma^{*} p \to p \rho^0$. The dashed curve shows the result of the GK calculation and the thin solid curve shows the result of the 
VGG calculation. The thick solid curve is the VGG calculation with the addition of the new $t$-channel meson exchange term. The dot-dashed curve shows the results of the
Regge JML calculation. The references of the presented data can be found in ref. \cite{rho}.}
\label{theoryRhzero}
\end{center}
\end{figure}

The Regge approach is most appropriate above the resonance region and at forward angles where the cross 
section is the largest. It consists of exchanges of meson ``trajectories'' in the $t$-channel.
In the following, we will use
   the ``JML'' acronym to refer to the specific model developed by J.-M. Laget and collaborators \cite{JML}.\\
For the $\rho^0$ channel, the model consists of $t$-channel exchanges of  $\sigma$, $f_2$ and Pomeron
trajectories. For the $\omega$ channel, the exchange particles are 
$\pi^0$ and Pomeron trajectories. For the $\phi$ channel, the only exchange particle is the Pomeron trajectory. 
For the $\rho^+$
channel, the exchange particles are the  $\pi^+$ and the $\rho^+$ trajectories. The free parameters are the coupling constants at the hadronic vertices
(most of them being well constrained) and the mass scales of the electromagnetic form factors.\\
Fig. \ref{theoryRhzero} shows the total longitudinal cross section  $\sigma_{L}(\gamma^* p \to p \rho^0)$ as a function of $W$ for fixed $Q^2$. The results of the  JML calculation 
is shown with the dot-dashed curve. The JML model is able to successfully reproduce the cross sections for
almost all of  our $(Q^2,W)$ range.
It can also reproduce successfully the main trends of the $W$, $Q^2$ and $t$ dependences of the total and differential cross
sections of the reactions $\gamma^* p \to p \omega$, $\gamma^* p \to p \phi$ and $\gamma^* p \to n \rho^+$ as presented in 
refs. \cite{omega,phi,mathese}. 
 \subsection{The GPD  ``partonic'' approach}
\begin{figure}[!h]
\begin{center}
\includegraphics[height=7cm,width=10cm]{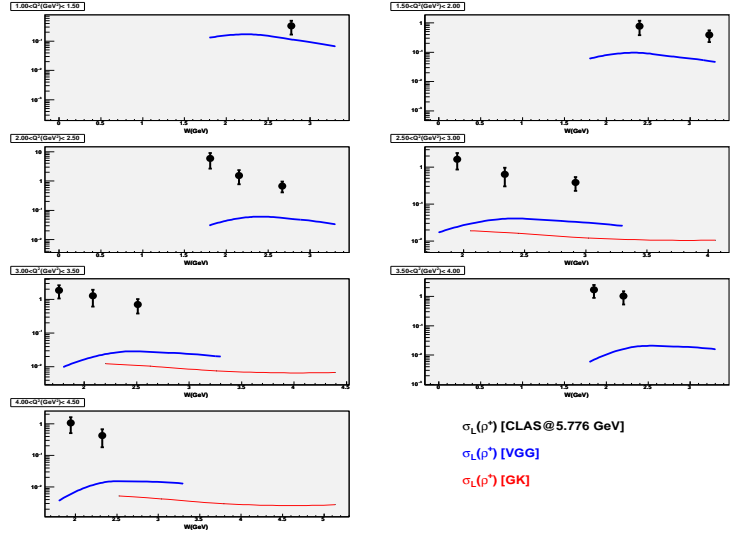}
\caption{\underline{PRELIMINARY} longitudinal cross section as a function of $W$ at fixed $Q^2$, for the reaction 
$\gamma^{*} p \to n \rho^+$. The red and the blue curves are the results of the GK and VGG models. Units are arbitrary on the y axis.}
\label{theoryRhoplus}
\end{center}
\end{figure}

\begin{figure}[!h]
\begin{center}
\includegraphics[height=5cm,width=8cm]{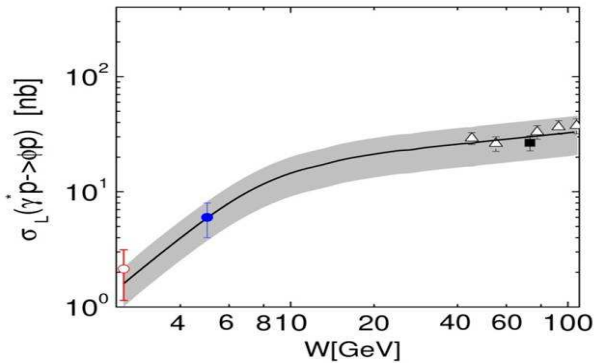}
\caption{Longitudinal cross section as a function of $W$, for the reaction 
$\gamma^{*} p \to p \phi$. The references of the presented data can be found in ref. \cite{phi}. The black curve is  
the result of the GK model.}
\label{phiGPDs}
\end{center}
\end{figure}

 The formalism of GPDs is valid in the so-called
Bjorken regime, i.e. $Q^2,\nu \to \infty$  with $x_B=\frac{Q^2}{2M\nu}$ finite. It was proven \cite{Collins97} that the dominant 
process for exclusive meson electroproduction, in the Bjorken limit, is given by
the so-called handbag diagram represented in Fig. \ref{reggeGPDs}. 
The handbag diagram is based on the notion of factorization in leading-order/leading-twist pQCD between a hard scattering process,
 exactly calculable in pQCD, and a nonperturbative nucleon structure part that is parametrized by the GPDs. For mesons, the factorization
  of the handbag diagram is only valid for the longitudinal part of the cross section.
For electroproduction of $\rho^0$ and $\omega$ we are sensitive to the sum of the quark handbag diagram and the
gluon handbag diagram, while for the  $\rho^+$ and the $\phi$ channels
we are sensitive to, respectively, only the quark handbag diagram and the gluon handbag diagram.\\
In the following  we discuss the two particular GK \cite{GK} and VGG \cite{VGG} GPD-based calculations that provide
quantitative results for the longitudinal part of the exclusive meson cross
sections. In Fig. \ref{theoryRhzero}, the  dashed line shows the results of the GK model, while the thin solid line shows the result of the VGG model. We see that they give a good description of the high and
intermediate $W$ region, down to $W \sim 5$ GeV. At high $W$ the slow rise of the cross section is due to the gluon and sea contributions, while the valence quarks contribute only at
small $W$. At lower $W$ values, where the new CLAS data lie, both the GK and VGG models fail to reproduce the data. This discrepancy can reach an order of magnitude at the lowest $W$
values. The trend of these particular GPD  calculations is to decrease as $W$ decreases, whereas the data increase. 
The same behavior was observed, in the low $W$ region, for the exclusive electroproduction of the
$\rho^+$ as one sees in Fig. \ref{theoryRhoplus}. The results of the calculations of the GK and VGG models are
  shown, respectively, with the red and the blue curves.\\
An attempt to reconcile the GPD calculation with the low $W$ $\rho^0$ cross sections is
presented in ref.\cite{mick}. Through a toy-model, $t$-channel meson exchanges
are included in the GPDs and the result of this calculation (actually a fit)
is illustrated by the thick blue curve in Fig. \ref{theoryRhzero}.\\
 Fig. \ref{phiGPDs} shows the longitudinal cross section of $\phi$
electroproduction  compared with the calculation of the GK model. One can see that the model reproduces well the data for
all the presented $W$ range: for low $W$ data where the CLAS data lie and for higher $W$. As mentionned before, for the $\phi$ channel, we have
only the contribution of the handbag diagram of gluons and this comparison shows that the GPD approach can work well for gluons.\\
From the results of the three mesons channels $\rho^0$, $\rho^+$ and $\phi$, one can conclude that the GPD approach seems to work for gluons
and sea quarks and largely fail for valence quarks. Many theorical efforts are now in progress to understand this transition between
the small $W$ region and the high $W$ region.\\  
 For the $\omega$ channel, SCHC was found to be violated by a significant amount, thus preventing a
longitudinal/transverse cross section separation. We therefore do not discuss here its comparison with
GPD calculations.
\subsection{Comparison of the $t$ slope for the $\rho^0$, $\omega$, $\phi$ and $\rho^+$ channels}
\begin{figure}[!h]
\begin{center}
\includegraphics[width=12cm]{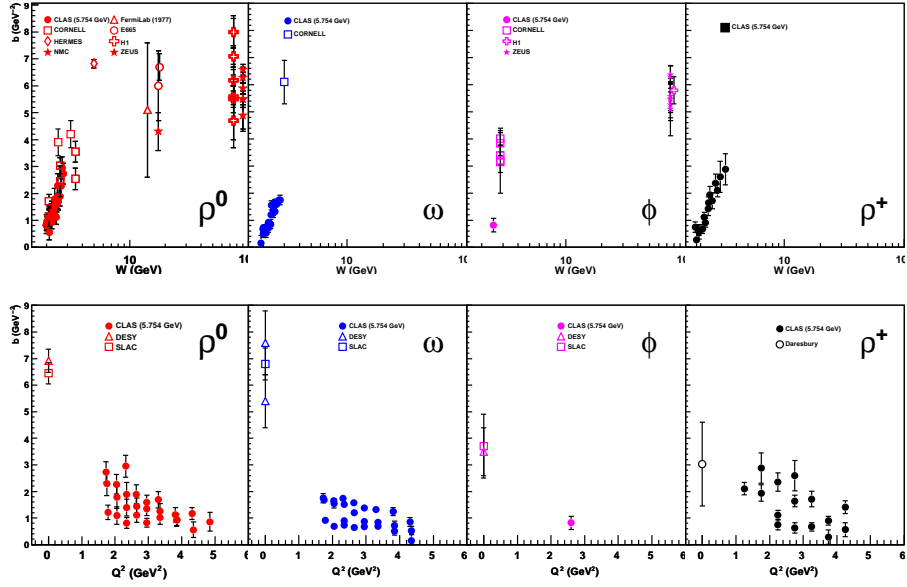}
\caption{The slope $b$ as a function of $W$ (on the top) and as a function of $Q^2$ (on the bottom) for the $\rho^0$, $\omega$, $\phi$
and $\rho^+$ channels.}
\label{tslopes}
\end{center}
\end{figure}

 Fig. \ref{tslopes} shows the slope of the differential cross section $d\sigma/dt$ for the  $\rho^0$, $\omega$, $\phi$ and $\rho^+$ channels as a function of $W$ (on the top part)
 and as a function of $Q^2$ (in the bottom part). One can see the same trends of this slope $b$ for all mesons channels, which can be interpreted
in simple and intuitive terms in the following way: 
 \begin{itemize}
 \item $b$ increases with $W$: the size of the nucleon increases as one probes the high $W$ values (i.e. the sea
 quarks), which could mean
 that the sea quarks tend to extend to the periphery of the nucleon.
 \item  $b$ decreases with $Q^2$: as we go to large $Q^2$, the resolution of the probe increases and we tend to see smaller and
 smaller objects.
\end{itemize}
\section{Summary}
Using the CLAS detector at JLab, we have collected the largest ever set of data for the exclusive electroproduction of vector mesons on the proton in the valence region. 
We have presented the published results  for  $\rho^0$, $\omega$ and $\phi$ and given a first look at the $\rho^+$ (preliminary) cross sections.
These data can be interpreted with two approaches:
 \begin{itemize}
   \item hadronic approach: the JML model describes well most of the features of the $\rho^0$, $\omega$, $\phi$ and $\rho^+$ cross
sections  up to $Q^2 \sim 4.5$ GeV$^2$.
 \item partonic approach: GPD models describe well the data for $W > 5$ GeV  but largely fail  for $W<5$ GeV. 
 \end{itemize}
A comparison between the features of the different mesons is in progress: we find for instance the same trends of the variation of the $t$
slope as a function of $W$ and as a function of $Q^2$ for all  $\rho^0$, $\omega$,
$\phi$ and $\rho^+$ channels. 

\end{document}